# Non-Negative Least Squares Reweighting and Pruning of Quadrature Grids for Tensor Hypercontraction

Andreas Erbs Hillers-Bendtsen[1,2], Lixin Lu[1,2], Todd J. Martínez[1,2,*]

[1]Department of Chemistry and The PULSE Institute, Stanford University, Stanford, CA 94305, USA

[2]SLAC National Accelerator Laboratory, 2575 Sand Hill Road, Menlo Park, CA 94025, USA

Email: Todd.Martinez@stanford.edu

**Abstract:** Tensor hypercontraction provides an attractive four-center two-electron repulsion integral format that can lower the scaling of many electronic structure methods while only requiring $O(N^2)$ memory. However, in its grid-based least-squares incarnation, tensor hypercontraction requires the tedious design of compact spatial quadrature grids to achieve efficiency and accuracy, representing a bottleneck for widespread application. To simplify grid generation, we devise a reweighting scheme in which the grid weights are optimized to ensure accurate reproduction of the atomic orbital overlap matrix by numerical integration. By casting this fitting task as a non-negative least-squares problem, we obtain a black-box methodology that not only yields robust grids for tensor hypercontraction as well as numerical integration of other integrals but also prunes the grids by zeroing quadrature weights for insignificant points.

**TOC GRAPHICS**

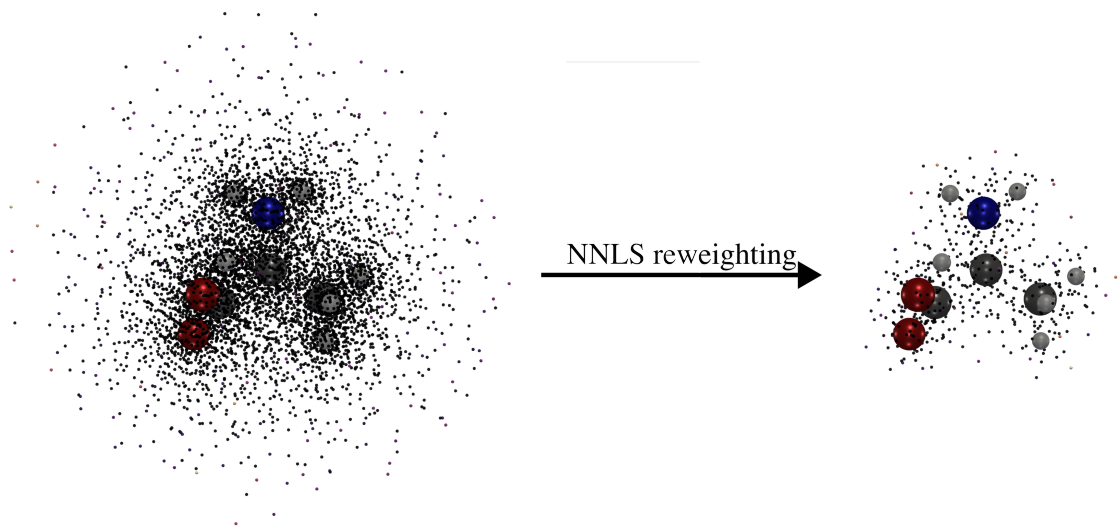

**Introduction**

In electronic structure theory, a central source of computational complexity lies in the calculation and manipulation of four-center two-electron repulsion integrals (ERIs) over Gaussian atomic orbitals $\phi(\vec{r})$:

$$(\mu\nu|\lambda\sigma) = \iiint d\vec{r}_1 \iiint d\vec{r}_2\ \phi_\mu(\vec{r}_1)\phi_\nu(\vec{r}_1)\frac{1}{r_{12}}\phi_\lambda(\vec{r}_2)\phi_\sigma(\vec{r}_2) \quad (1)$$

Fundamentally, ERIs are required for all electronic structure methods, ranging from Hartree-Fock theory to coupled cluster (CC) theory. The main challenge in handling ERIs is their formal $O(N^4)$ computational scaling and, in particular, the $O(N^4)$ memory requirements, necessitating the development of efficient techniques to avoid explicit storage of the full four index tensor. Most commonly, the problem is addressed in a brute-force manner by repeatedly recalculating the ERIs using efficient algorithms[1-3] and contracting them with other quantities on the fly whenever they are required. Depending on the electronic structure method, various specialized algorithms exist for directly computing the contraction of some quantity with the ERIs, thereby avoiding explicit storage.[4-14] These specialized approaches exploit sparsity in the ERI tensor to avoid calculating entries that can be predicted to be negligible and they can consequently achieve $O(N^2)$ or even $O(N)$ scaling in some contexts. Although highly efficient ERI evaluation algorithms exist, the cost of such integral-direct approaches can become substantial, ultimately limiting both the run time of a calculation and the system sizes that can be considered.

Another approach to circumvent the $O(N^4)$ memory requirements is to decompose the ERIs into products of smaller tensors or matrices. Within this class of methods, density fitting (DF) (also known as the resolution of the identity approximation)[15] and Cholesky decomposition (CD)[16,17] remain the most widely used approaches. Both decompose the ERIs into a product of two three-index tensors as

$$(\mu\nu|\lambda\sigma) \approx \sum_A B^A_{\mu\nu} B^A_{\lambda\sigma} \quad (2)$$

where $A$ is a set of auxiliary functions whose form depends on the chosen decomposition technique. Effectively, DF and CD both reduce the memory requirements of the ERIs to $O(N^3)$, permitting explicit storage for up to roughly 3000 basis functions using 1 TB of memory. Coupled with their increased mathematical flexibility, DF and CD therefore enable a significant reduction in the computational prefactor associated with many electronic structure methods including, but not limited to, self-consistent field (SCF) methods[18, 19] such as density functional theory (DFT), Møller-Plesset perturbation theory,[20, 21] and CC theory.[22] However, DF and CD are unable to effectively lower the formal scaling of these methods as they do not fully decouple all orbital indices, and the $O(N^3)$ storage requirements limit the reachable system sizes. Another relevant approach to mention is pseudospectral methods.[23-30] Unlike DF and CD, pseudospectral methods can lower the scaling of exchange-like contributions by fully decoupling two of the four orbital indices. However, they are limited by the large quadrature grids used to high ensure accuracy and by their violation of the permutational symmetry of the ERIs.

The most effective of ERI decomposition to date is the more recent tensor hypercontraction (THC) scheme[31-34] which approximates the ERI tensor as the product of five matrices

$$(\mu\nu|\lambda\sigma) \approx \sum_{PQ} X_\mu^P X_\nu^P Z^{PQ} X_\lambda^Q X_\sigma^Q \qquad (3)$$

effectively reducing the storage requirements to $O(N^2)$, assuming that the number of components $P$ and $Q$ grows linearly with system size $N$. Furthermore, THC unpins all four atomic orbital indices, providing increased mathematical flexibility that enables lowered scaling implementations of many electronic structure methods,[31, 32, 35-39] for example formally $O(N^3)$ SCF methods.[40, 41] The primary challenge in applying THC is the construction of the decomposition itself. The *de facto* standard THC construction scheme is grid-based least-squares THC (LS-THC). In LS-THC, the $X$ matrices are predefined as weighted collocation matrices on a spatial quadrature grid and the central $Z$ metric matrix is then determined by minimizing the least-squares error in the reproduced ERIs.[33] Using LS-THC, construction of Z scales as $O(N^5)$ with conventional four-index ERIs and as $O(N^4)$ if the DF approximation is invoked. More recently, it has also been shown that an $O(N^3)$ scaling LS-THC construction can be obtained by fitting to two-center two-electron integrals in an auxiliary basis, without severely compromising the accuracy of the resulting THC decomposition.[40, 41] It should be noted that $O(N^3)$ scaling LS-THC construction has also been realized for periodic systems using plane-wave basis sets with periodic boundary conditions.[42-45]

A central component in the LS-THC approach is the spatial quadrature grid, since LS-THC can essentially be viewed as an implicit numerical quadrature of the ERIs. The grid is critical for obtaining an accurate THC decomposition, but it is also desirable for it to be as compact as possible to reduce both storage requirements and cost of subsequent computations. Previous studies have therefore focused on carefully optimizing grids for LS-THC using standard spherical grids combined with a radial discrete variable representation for which the radial nodes and associated weights are optimized.[46, 47] This approach is very tedious, as it requires optimization of grids for every element and basis set combination to be effective. Several studies have therefore explored effective grid point selection schemes, such as interpolative separable density fitting,[42] which has seen widespread success in periodic calculations, and the similar pivoted Cholesky decomposition based pruning of Matthews,[48] which can efficiently select grid points in a "black-box" manner from an overly large input grid. However, none of these approaches consider reweighting the pruned quadrature grid to further minimize the THC error and to make the quadrature grid useful for numerically integrating e.g. one electron integrals.

In this paper, we present a strategy for reweighting spatial quadrature grids for LS-THC by minimizing the error in the reproduction of the atomic orbital overlap matrix on the grid subject to a non-negativity constraint. We demonstrate that this strategy improves both the accuracy of the resulting LS-THC decomposition and the compactness of the grid by automatically setting insignificant weights to zero. Initially, we briefly outline the theory underlying LS-THC and the proposed grid reweighting scheme. We then present a series of numerical illustrations that highlight the performance of the grid reweighting scheme compared to the Cholesky based pruning method of Matthews,[48] in terms of accuracy, grid pruning, and the resulting computational efficiency gains in LS-THC-CASPT2 calculations. Finally, we give concluding remarks and discuss the potential application of the developed scheme in future studies.

**Theory**

In the LS-THC approach,[33] the THC decomposition in Eq. (3) is constructed via an implicit numerical quadrature on a real-space grid by minimizing the cost function

$$O_{LS-THC} = \frac{1}{2} \sum_{\mu\nu\lambda\sigma} ||(\mu\nu|\lambda\sigma) - \sum_{PQ} X_\mu^P X_\nu^P Z^{PQ} X_\lambda^Q X_\sigma^Q ||^2 \tag{4}$$

Here, $O_{LS-THC}$ represents the Frobenius norm of the error imposed by the THC approximation. In this approach, the $X$ matrices are defined as weighted collocation matrices

$$X_\mu^P = \phi_\mu(\vec{r}_P)(w_P)^{\frac{1}{4}} \tag{5}$$

where the quadrature grid has points $\vec{r}_P$ and associated weights $w_P$. The quadrature weights are typically computed using the Becke weighting scheme[49] developed for DFT quadrature grids and these are therefore not optimized for LS-THC. Once the $X$ matrices are constructed, the metric matrix $Z$ is chosen to minimize the cost function $O_{LS-THC}$ as:

$$Z^{PQ} = \sum_{RS} \sum_{\mu\nu\lambda\sigma} (M^{-1})_{PR} X_\mu^R X_\nu^R (\mu\nu|\lambda\sigma) X_\lambda^S X_\sigma^S (M^{-1})_{SQ} \tag{6}$$

where the grid metric matrix $M$ is defined as

$$M_{PQ} = \sum_{\mu\nu} X_\mu^P X_\mu^Q X_\nu^P X_\nu^Q \tag{7}$$

Computing the LS-THC decomposition via Eq. (6) thus scales as $O(N^5)$ with conventional ERIs, or $O(N^4)$ if the DF approximation is also invoked. Notably, in both cases, the asymptotically rate-limiting step scales linearly with the number of grid points, while the inversion of $M$, typically performed using a pseudoinverse, scales cubically with the number of grid points.

In LS-THC, the $X$ matrices can be interpreted as representing the atomic orbitals on the grid, while $X_\mu^P X_\nu^P$ represents the generalized charge density $\rho_{\mu\nu} = \phi_\mu(\vec{r})\phi_\nu(\vec{r})$ on the grid. The metric matrix $Z$ can therefore be viewed as an implicit encoding of the Coulomb repulsion between electrons in those orbitals. We can consequently assert that a fundamental requirement for a weighted quadrature grid in LS-THC is that it accurately reproduces the atomic orbital overlap matrix $S$ via numerical integration

$$S_{\mu\nu} = \iiint d\vec{r}_1 \iiint d\vec{r}_2 \, \phi_\mu(\vec{r}_1)\phi_\nu(\vec{r}_2) \approx \sum_P X_\mu^P \, w_P^{\frac{1}{2}} X_\nu^P = \sum_P \phi_\mu(\vec{r}_P) \, \phi_\nu(\vec{r}_P) w_P \qquad (8)$$

Standard quadrature grids used in quantum chemistry, such as Becke grids,[49] employ tabulated grid weights that are not specifically optimized to reproduce the overlap matrix $S$ for a given molecular system. A new set of weights that ensure a better reproduction of $S$, and likely also other molecular integrals, can be determined by minimizing the cost function

$$O_S = \frac{1}{2} \sum_{\mu\nu} ||S_{\mu\nu} - \sum_P \phi_\mu(\vec{r}_P) \, \phi_\nu(\vec{r}_P) w_P ||^2 \qquad (9)$$

with respect to the weights $w_P$. By defining $A_{\mu\nu}^P = \phi_\mu(\vec{r}_P) \, \phi_\nu(\vec{r}_P)$ and flattening the $\mu\nu$ indices into a compound index, the optimal $w$ is found from a linear system of equations

$$Aw = S \qquad (10)$$

which can be solved using standard linear algebra techniques, such as a least squares procedure. Since the grid and associated weights are used for numerical quadrature, the weights represent the volumes that are being integrated over for the given grid points. It is therefore only physically meaningful to work with non-negative weights, which can be achieved using a non-negative least squares (NNLS) algorithm, such as the Lawson–Hanson method.[50] In addition to enforcing non-negativity, the NNLS solution achieves two important goals: 1) The NNLS procedure may project many of the weights to zero, effectively pruning unnecessary points from the input quadrature grid. 2) The resulting weights are normalized, i.e. $\sum_P w_P = 1$, meaning that the output grid can still be used for numerical integration of other quantities.

Below, we compare the NNLS scheme to the pivoted Cholesky pruning procedure developed by Matthews[48] and we therefore briefly summarize the Cholesky procedure here. The pivoted Cholesky pruning approach prunes the input grid by discarding grid points using an incomplete pivoted Cholesky decomposition of the grid metric matrix $M$, where only columns with a pivot above a certain threshold are retained. This corresponds to keeping only grid points for which $\sum_{\mu\nu} X_\mu^P X_\mu^Q X_\nu^P X_\nu^Q$ is non-negligible, since these points will otherwise have a vanishing contribution to the ERIs that can be reconstructed from the resulting LS-THC decomposition.

While this approach is simple and computationally efficient, it does not preserve normalization of grid weights meaning that it can only be used for approaches such as LS-THC, where the weights are internally renormalized as $Z$ is fit. Another limitation is that the output grid can never achieve a better accuracy than the input grid.

In the following, we illustrate the advantages of NNLS reweighting of quadrature grids for LS-THC, focusing on both the resulting accuracy and the extent to which grid points can be pruned, thereby reducing storage requirements and the computational cost of subsequent calculations.

**Numerical Results**

*Computational Details*

A pilot implementation of the NNLS grid reweighting was developed in Python, with PySCF[51] as a backend for obtaining the necessary molecular integrals. The NNLS solution to Eq. (10) was computed using an active set method based on the Lawson–Hanson algorithm,[52] with convergence achieved when all weights were either passive or their corresponding gradient components fell below a predefined threshold, which we refer to as the the weight threshold below. This corresponds to the Karush-Kuhn-Tucker conditions for NNLS problems. After NNLS convergence, we discard all points that have an associated weight of zero, as their contributions vanish. We use either the cc-pVDZ-rdvr3 and cc-pVTZ-rdvr3 hand-optimized grids[47] for the cc-pVDZ and cc-pVTZ basis sets implemented in the TeraChem program[53, 54] or DFT quadrature grids from PySCF as input grids for NNLS reweighting. For comparison, we also prune the grids using the pivoted Cholesky procedure retaining only pivots and their associated grid points above the specified weight threshold. For subsequent complete active space second order perturbation theory (CASPT2) calculations, the reweighted grids are imported into TeraChem, and the existing LS-THC-CASPT2 implementation[37] is used. For comparison to standard CASPT2, we compare to results obtained using BAGEL.[55]

Below, we show the results from a series of test calculations to investigate the NNLS reweighting. First, we examine the characteristics of the NNLS reweighting using alanine as a test case and compare it to the pivoted Cholesky pruning. For these calculations, we quantify the error in reproducing the AO overlap matrix using numerical integration on the grid and the error of the LS-THC approximation of the ERIs using the root mean square deviation (RMSD). We compute the RMSD for $S$ and for $(\mu\nu|\lambda\sigma)$ as

$$\text{RMSD for } S = \frac{O_S}{\sqrt{n_{AO}^2}} \qquad (11)$$

and

$$\text{RMSD for } (\mu\nu|\lambda\sigma) = \frac{O_{LS-THC}}{\sqrt{n_{AO}^4}} \qquad (12)$$

respectively, where $O_S$ and $O_{LS-THC}$ are the Frobenius errors given in Eqs. (9) and (4), respectively, while $n_{AO}$ is the number of atomic orbitals. The RMSDs gives a statistical estimate of the error in each entry in the respective tensor. Following the analysis with alanine, we study the computational savings and accuracy obtained from NNLS reweighted grids in CASPT2 calculations.

*Analysis of grid reweighting for alanine*

In Figure 1, we display results for alanine in the cc-pVDZ basis set obtained after NNLS reweighting or pivoted Cholesky pruning of the smallest DFT level 0 grid from PySCF with varying weight thresholds. More specifically, we show the RMSD between the analytical $S$ and the approximation to it from numerical integration on the grid, the RMSD between the analytical ERIs $(\mu\nu|\lambda\sigma)$ and those obtained for the resulting LS-THC decomposition, the number of grid points retained after pruning, and the (4,4)-LS-THC-CASPT2 energy obtained using the pruned grids.

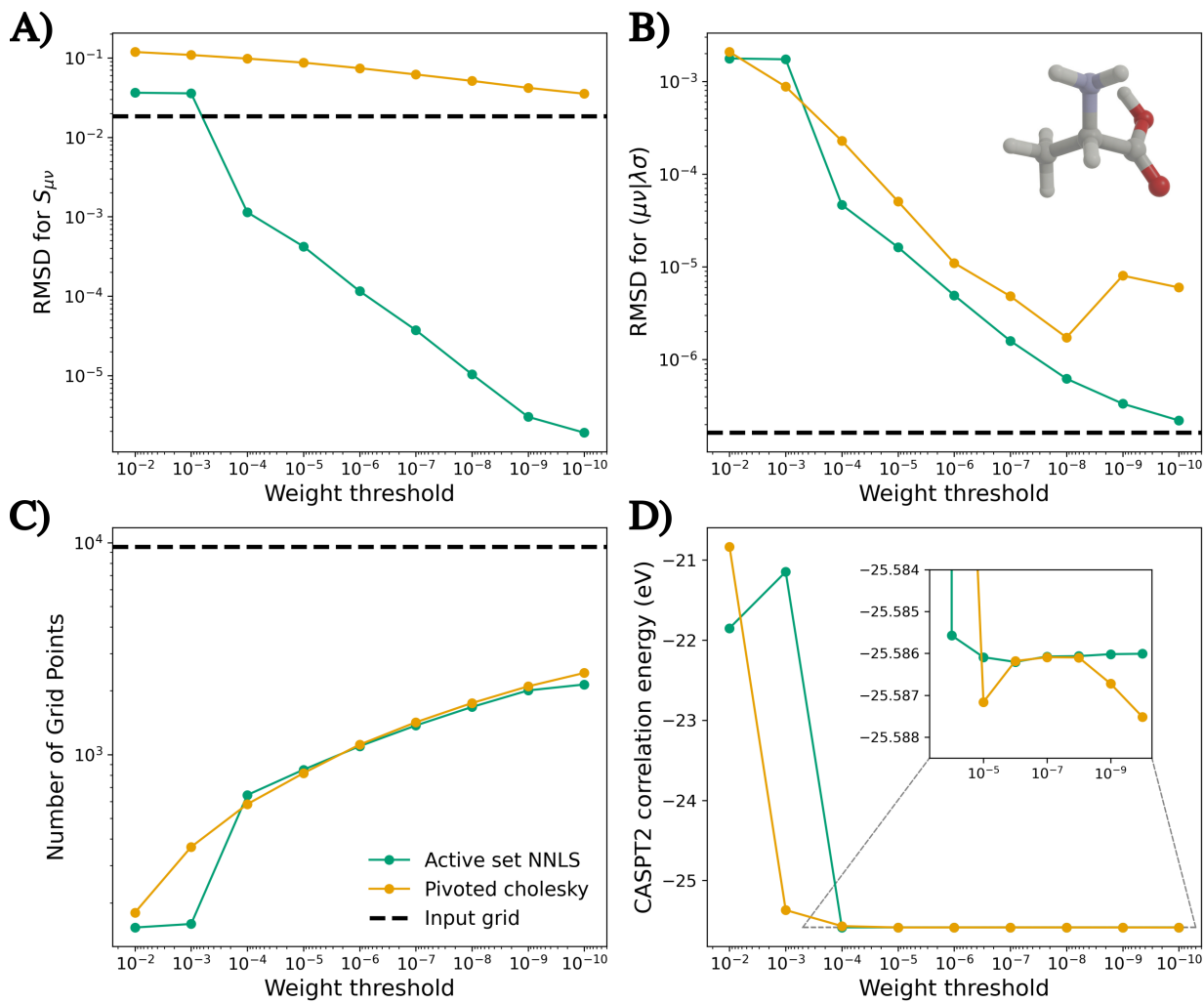

**Figure 1**. Analysis of weight threshold in the NNLS reweighting according to Eq. (10) and the pivoted Cholesky pruning for alanine in a cc-pVDZ basis set using the level 0 DFT grid from PySCF as input grid. A) RMSD for $S$. B) RMSD for $(\mu\nu|\lambda\sigma)$ from LS-THC. C) Number of grid points after NNLS reweighting or Cholesky pruning. D) (4,4)-LS-THC-CASPT2 correlation energies. Panels A)–C) are plotted on double logarithmic scales while panel D) only has a logarithmic x-scale.

Figure 1 shows that the DFT level 0 grid from PySCF provides a robust LS-THC fit, since the grid is very large in the context of LS-THC and therefore leads to limited compression relative to the original ERI tensor. NNLS reweighting improves reproduction of $S$ compared to the input grid with a convergence threshold of $10^{-4}$ or tighter. Similarly, the corresponding RMSD for the ERIs steadily drops with tighter thresholds and reaches the same accuracy as the input grid for a threshold of $10^{-10}$. For all thresholds studied, the NNLS procedure discards a significant portion of grid points, roughly 80% for $10^{-10}$ convergence threshold, as the associated weights are set to zero (going from a grid with 735 points/atom to a pruned grid with 165 points/atom). This indicates that the input grid is, as expected, much larger than needed. Furthermore, fewer grid points are retained at looser convergence thresholds, since the NNLS procedure will zero weights that are predicted to be below the threshold. Interestingly, both the NNLS procedure and the pivoted Cholesky pruning find similar grid sizes. This suggests that both methods select the most important points from the input grid with comparable precision. Meanwhile, the NNLS procedure yields a more accurate LS-THC fit, since the weights are refitted to properly reproduce the overlap matrix, whereas the pivoted Cholesky pruned grids show poorer overlap matrix (and ERI) reproduction. This further results in quicker and more stable convergence on the (4,4)-LS-THC-CASPT2 correlation energies for NNLS, where the change in energy upon tightening the threshold further drops below 1 $\mu$eV at $10^{-4}$. Correspondingly, the happens at $10^{-6}$ for pivoted Cholesky pruning that consequently requires larger grids to achieve the same precision in the resulting energies. Based on these results, we use a weight threshold of $10^{-4}$ for NNLS and $10^{-6}$ for pivoted Cholesky pruning in all of the following calculations, since this threshold provides a reasonable balance between LS-THC accuracy and grid compactness, while avoiding the risk of overfitting to $S$ in the NNLS procedure.

We now examine the dependence on the input grid for the two procedures by displaying the number of grid points and the resulting RMSD for the $(\mu\nu|\lambda\sigma)$ resulting from the LS-THC fit for each grid in Figure 2. It should be noted that the DFT 2 and DFT 3 grids are so large that the pseudoinversion of $M$ could not be performed using SciPy due to integer overflow, however, the THC decomposition resulting from these would essentially be exact.

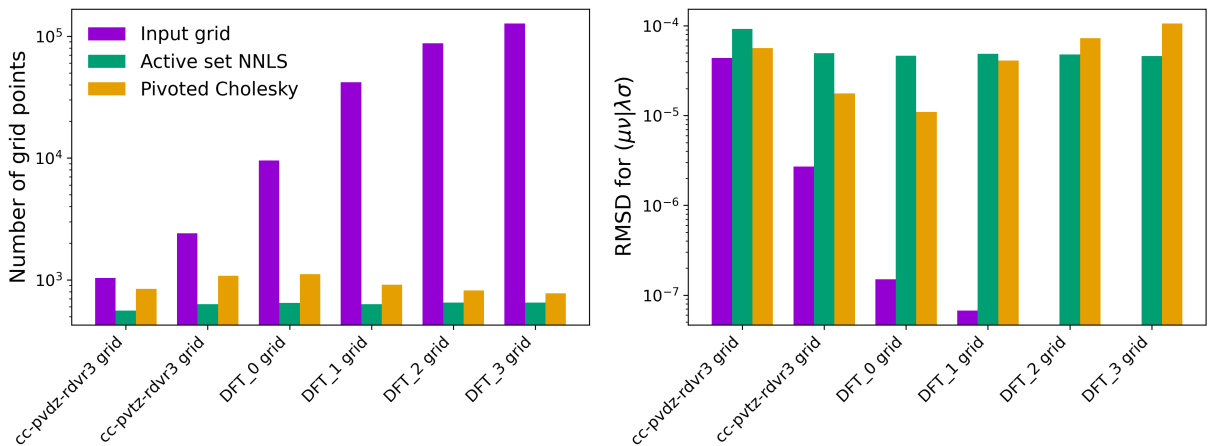

**Figure 2**. Grid size obtained for alanine in a cc-pVDZ basis set using NNLS reweighting according to Eq. (10) with a weight threshold of $10^{-4}$ and the pivoted Cholesky pruning with a weight threshold of $10^{-6}$. Left: Number of grid points before and after pruning. Right: RMSD for $(\mu\nu|\lambda\sigma)$ from LS-THC before and after pruning and/or reweighting.

Firstly, both the NNLS reweighting and pivoted Cholesky procedures are very effective at pruning the grid. Secondly, the NNLS reweighted grids provide a comparable RMSD for the ERIs to that obtained with the cc-pVDZ-rdvr3 grid, which was specifically optimized for the cc-pVDZ basis set, but it achieves this accuracy with significantly smaller grids. Meanwhile, the pivoted Cholesky procedure generally produces similar errors but it does so with significantly larger grids, as the weights are not refit. Interestingly, the number of grid points retained by both procedures naturally depends on the size of the input grid up to a certain point. For the NNLS reweighting, the number of retained grid points does not increase significantly beyond the DFT level 0 grid, and the accuracy of the resulting THC decomposition does not improve significantly. This indicates that the space in which the NNLS procedure can select points is saturated, so increasing the grid size offers no additional benefit at the given weight threshold. For the pivoted Cholesky pruning, the output grid size actually decreases for the very large DFT grid leading to an increase in the RMSD for the ERIs. This occurs because the weight associated with each grid point decreases for bigger grids, causing the pivots of $M$ to be smaller, so a greater fraction of the points is discarded. While both methods efficiently prune the input grids, the NNLS procedure yields grids that are significantly more compact while consistently yielding a robust LS-THC fit with acceptable accuracy. This is even the case for the cc-pVDZ-rdvr3 grid which is specifically optimized for the cc-pVDZ basis set.

The grid dependence naturally also varies with the basis set and we therefore assess the dependence on the basis set for alanine using the DFT level 0 grid with various double-$\zeta$ and triple-$\zeta$ basis sets.

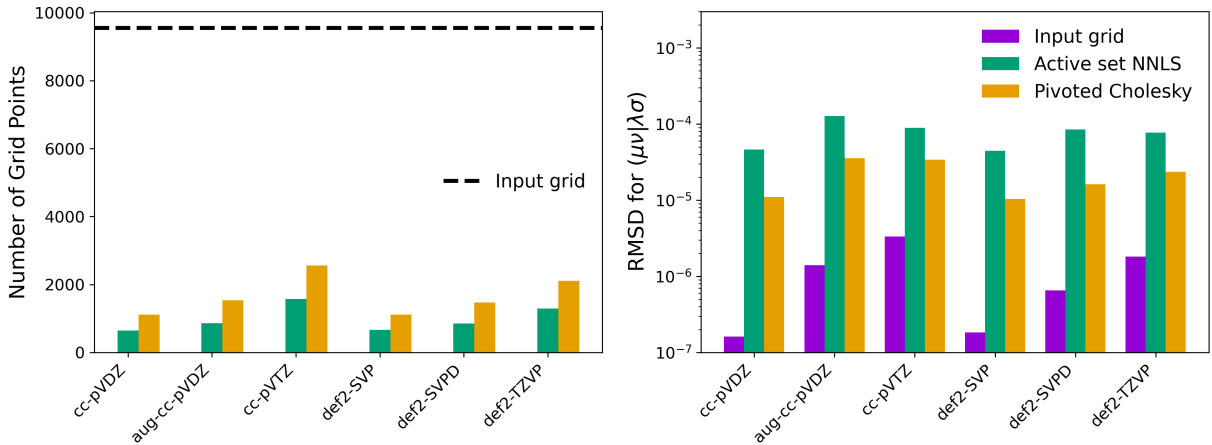

**Figure 3.** NNLS reweighting according to Eq. (10) and pivoted Cholesky pruning for alanine in various basis sets using the DFT level 0 grid from PySCF, a NNLS weight threshold of $10^{-4}$, and a pivoted Cholesky pruning weight threshold of $10^{-6}$. Left: Number of grid points after NNLS reweighting or pivoted Cholesky pruning. Right: RMSD for $(\mu\nu|\lambda\sigma)$ from LS-THC before and after NNLS reweighting or pivoted Cholesky pruning.

Figure 3 clearly shows that the NNLS procedure can optimize the grid for different basis sets, thus providing a compact grid tailored to the specific combination of system and basis set. Unsurprisingly, larger basis sets require larger grids, yet, the ratio between number of grid points and number of basis functions remain relatively constant within a specific basis set family (See Table S1). We also note that the RMSD for the ERIs also grows slightly as the basis set size increases since the input grid is unchanged. Nevertheless, the NNLS optimized grids are significantly smaller and provide useful RMSDs in all cases. We expect this trend to hold across other basis sets, provided the input grid is sufficiently large for the specific basis set. For the pivoted Cholesky pruning, the output grids are again larger than the NNLS and they therefore generally exhibit slightly lower RMSDs for the ERIs.

*Accuracy and computational speedup in LS-THC-CASPT2 calculations*

To assess the effect of using NNLS optimized grids and pivoted Cholesky pruned grids on both wall time and accuracy in actual applications of LS-THC, we performed a series of CASPT2 calculations. We examine the accuracy of LS-THC-CASPT2 for relative energy profiles of Diels-Alder reactions between *cis*- or *trans*-butadiene and ethylene using structures from a previous study[37] that model both the concerted (con) and two-step (bi) reaction pathways. For these calculations, we use a (6,6) active space and the cc-pVDZ-rdvr3 and cc-pVTZ-rdvr3 grids from TeraChem, as well as the grids that result from running either NNLS optimization or pivoted Cholesky pruning of these.

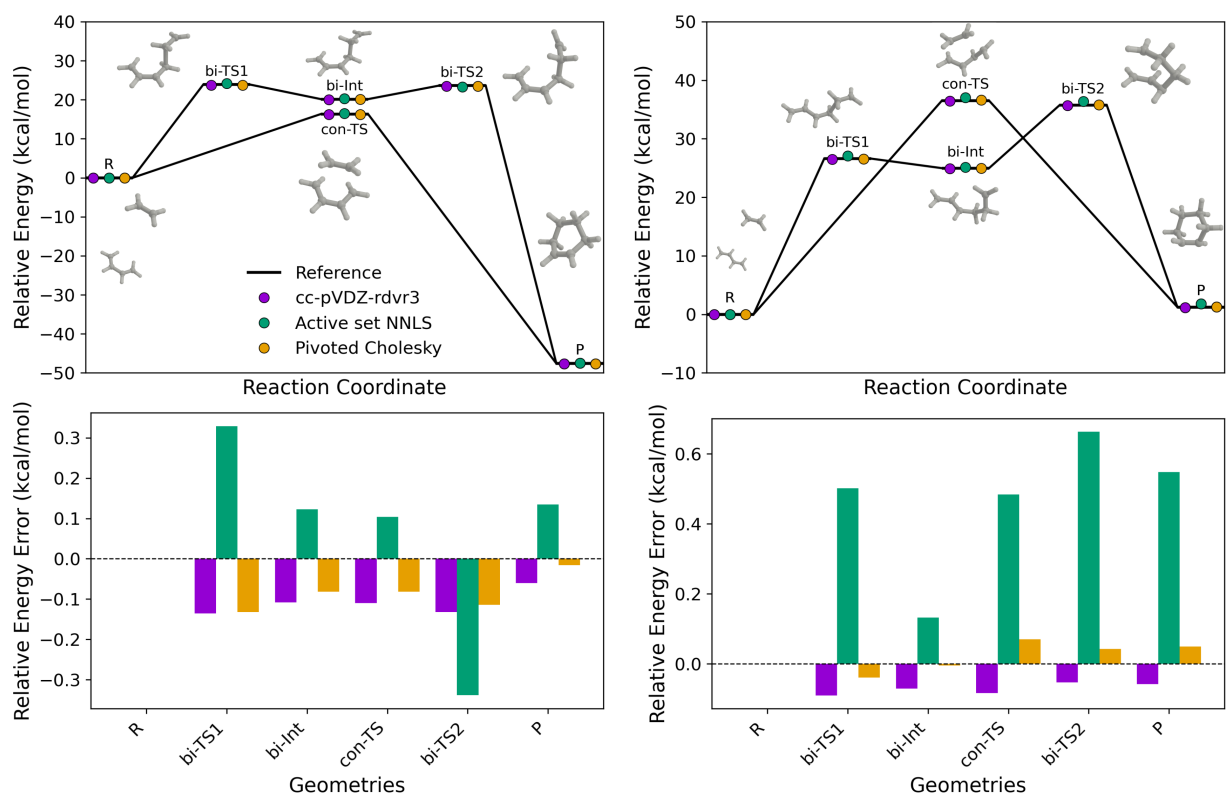

**Figure 4.** Diels-Alder reaction profile for cis/trans-butadiene starting from the cc-pVDZ-rdvr3 grid: (Top) Relative CASPT2 energies for the different geometries with cis-butadiene on the left and trans-butadiene on the right. (Bottom) Relative energy error for THC-CASPT2 compared to conventional CASPT2 using different quadrature grids with cis-butadiene on the left and trans-butadiene on the right.

Figure 4 and Figure 5 show that the relative energy profiles from conventional CASPT2 and LS-THC-CASPT2 are qualitatively indistinguishable for both *cis*- and *trans*-butadiene for the native cc-pVDZ-rdvr3 and cc-pVTZ-rdvr3 grids from TeraChem with the errors being dominated by the underlying density fitting error. For the native TeraChem grids, LS-THC errors in the relative CASPT2 energies are smaller than 0.20 kcal/mol indicating that LS-THC does not introduce significant errors that could lead to misinterpretation of the reaction profiles. Upon pruning the cc-pVDZ-rdvr3 and cc-pVTZ-rdvr3 grids, minor differences are seen in the energy profiles and the relative energy error changes for both NNLS and pivoted Cholesky pruning. Comparing different grids, the NNLS and pivoted Cholesky pruned grids maintain accuracy within 0.6 kcal/mol of the reference CASPT2 values and still provide the same conclusion as the results from the cc-pVDZ-rdvr3 and cc-pVTZ-rdvr3 grids.

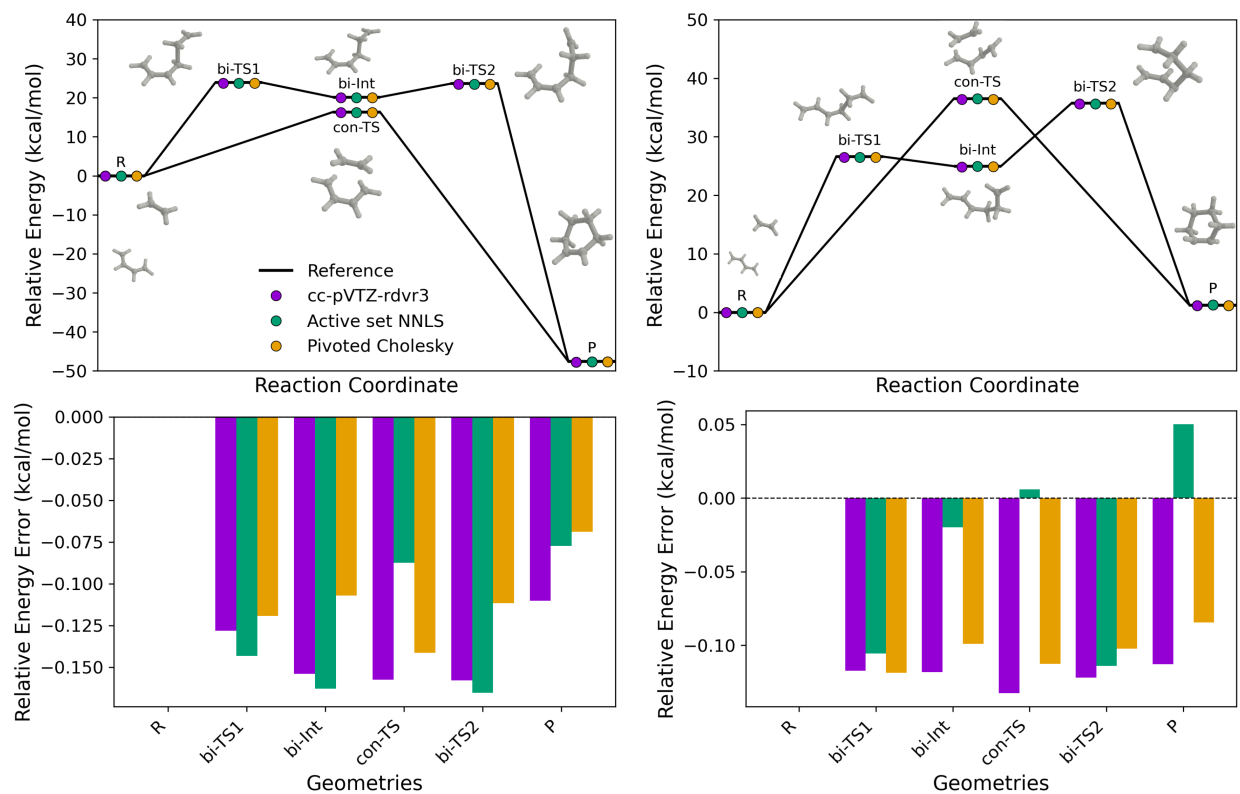

**Figure 5.** Diels-Alder reaction profile for cis/trans-butadiene starting from the cc-pVTZ-rdvr3 grid: (Top) Relative CASPT2 energies for the different geometries with cis-butadiene on the left and trans-butadiene on the right. (Bottom) Relative energy error for THC-CASPT2 compared to conventional CASPT2 using different quadrature grids with cis-butadiene on the left and trans-butadiene on the right.

This is achieved with substantially smaller grid sizes, particularly for NNLS with the cc-pVDZ-rdvr3 input grid, where an average of only 608 out of approximately 1185 input grid points are retained (see Table 1). In general, the pivoted Cholesky procedure leads to slightly larger grids than the NNLS procedure and its errors therefore appear to fluctuate less in sign and magnitude overall compared to NNLS. The NNLS procedure provides much smaller grids without significantly increasing the numerical error in the energy. However, the NNLS pruned grids are so small that the errors are not dominated by the underlying density fitting error. Instead, it is a combination of the THC and density fitting errors which cancel to different extents for different geometries and grids. This is what causes the fluctuations in sign and magnitude of the errors of the both NNLS and pivoted Cholesky pruned grids. Comparing the two input grids, using the larger cc-pVTZ-rdvr3 grid does lead to pruned grids that give slightly smaller errors but also slightly larger grids compared to starting from the smaller cc-pVDZ-rdvr3 grid. However, the NNLS grids obtained by pruning the cc-pVTZ-rdvr3 generally leads to relative energy errors that fluctuate less and align more with those obtained from the input grid indicating that the THC error is smaller than density fitting error which then dominates.

| Basis Set | Input Grid | Active Set NNLS | Pivoted Cholesky |
|---|---|---|---|
| cc-pVDZ-rdvr3 | 1185 | 608 | 944 |
| cc-pVTZ-rdvr3 | 2872 | 680 | 1189 |

**Table 1.** Average number of grid points used for the THC-CASPT2 calculation of the Diels-Alder reactions for different grid types.

To evaluate the computational speedup achieved in LS-THC-CASPT2 using smaller grids obtained via NNLS optimization and pivoted Cholesky pruning, we compute the ground-state energies of butadiene and hexatriene solvated in an increasing number of methanol molecules, using geometries from a previous study.[37] For butadiene and hexatriene, we use active spaces of (4,4) and (6,6), respectively, and in both cases with start from a cc-pVDZ-rdvr3 grid, which is then used as-is or pruned.

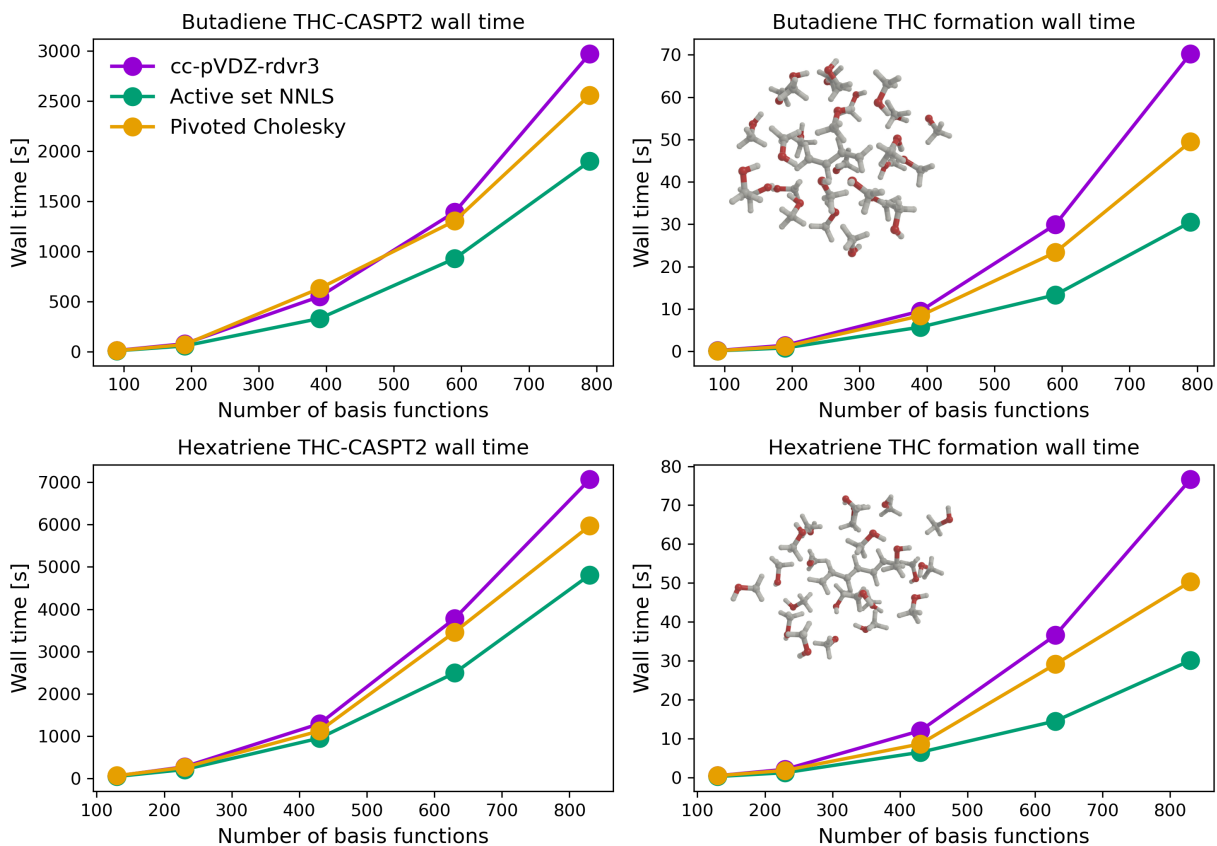

**Figure 6.** Wall times for LS-THC-CASPT2 and THC formation with different grids for butadiene and hexatriene with an increasing number of methanol solvent molecules. The LS-THC-CASPT2 wall time includes the time spent on the THC formation and computing PT2 correction (all iterations), but it does not include the preceding CASSCF calculation.

Figure 6 shows the smaller grids obtained from NNLS optimization and pivoted Cholesky pruning naturally lead to reduced wall times for both the overall THC-CASPT2 calculation and the THC formation step across all systems. For the largest systems containing 14 methanol molecules, THC formation using the NNLS optimized grid is more than twice as fast as using the input cc-pVDZ-rdvr3 grid for both butadiene and hexatriene. Furthermore, the corresponding total THC-CASPT2 wall times are around 25-35% lower, indicating that grid pruning and optimization can lead to significant time savings. Meanwhile, the grids obtained from pivoted Cholesky pruning offer smaller but still substantial wall time reductions than NNLS. This is primarily because the grids are larger than those obtained via NNLS reweighting as seen from Table S2. Nevertheless, it is evident that the savings in wall time are roughly proportional to the grid reduction, meaning that substantial wall time savings can be achieved through efficient grid pruning.

Finally, we note that we have not included the grid optimization time in the timings, since our current pilot implementation of the active set NNLS grid optimization is a naive, unoptimized prototype and not fast enough to allow meaningful comparison of wall times. The pivoted Cholesky pruning, on the other hand, is quite efficient out-of-the-box using LAPACK routines and does not add extra wall time to THC construction. In future studies, we aim to devise an efficient black-box grid optimization procedure that can both prune and reweight input grids.

**Conclusion**

In this paper, we have outlined a procedure to efficiently prune and reweight spatial quadrature grids for LS-THC and potentially other numerical quadratures in quantum chemistry. The quadrature weights are determined via a NNLS fit to the atomic orbital overlap matrix to ensure that the grid is tailored to accurately model the generalized charge density arising from products of primary basis functions. Following this approach, we achieve a robust THC decomposition with a compact grid as the NNLS procedure automatically sets redundant weights to zero, yielding dramatically smaller grids that provide THC errors comparable to much larger input grids. The procedure readily extends to arbitrary choices of basis set and input grid combinations providing grids tailored to the specific molecule/basis set combination. Compared to similar grid pruning methods, the developed NNLS reweighting provides similarly small grids, but, importantly, it also allows for a refitting of the quadrature weights making it more accurate and robust across different input grids. Using the resulting grids in LS-THC-CASPT2 calculations, we achieve significant speedups compared to similar calculations using carefully optimized grids from previous studies with minimal loss of accuracy. This benefit naturally extends to any method that relies on LS-THC, since the number of grid points directly affects the scaling of such methods.

In future studies, we will pursue an efficient implementation of the NNLS reweighting scheme to facilitate on-the-fly optimization of grids for LS-THC. This will enable robust black-box determination of quadrature grids for any molecule in any basis set. Effectively, this will circumvent the need for tedious optimization of quadrature grids for every single element/basis set combination and make widespread adoption of LS-THC easier.

## Supporting Information

Basis set and grid information. All data and python scripts for data generation and treatment and figure generation are given on Zenodo (https://doi.org/10.5281/zenodo.19356231).

## Notes

TJM is a founder of PetaChem, LLC.

## Acknowledgements

This work was supported by the AMOS program of the U. S. Department of Energy, Office of Science, Basic Energy Sciences, Chemical Sciences and Biosciences Division. AEHB acknowledges financial support from the Novo Nordisk Foundation under grant reference number NNF24OC0089345.

## References

1. McMurchie, L. E.; Davidson, E. R., One-and two-electron integrals over Cartesian Gaussian functions. *J. Comp. Phys.* **1978,** *26*, 218.
2. Rys, J.; Dupuis, M.; King, H., Computation of electron repulsion integrals using the Rys quadrature method. *J. Comp. Chem.* **1983,** *4*, 154.
3. Obara, S.; Saika, A., Efficient recursive computation of molecular integrals over Cartesian Gaussian functions. *J. Chem. Phys.* **1986,** *84*, 3963.
4. Almlöf, J.; Fægri Jr, K.; Korsell, K., Principles for a direct SCF approach to LCAO–MO ab-initio calculations. *J. Comp. Chem.* **1982,** *3*, 385.
5. Häser, M.; Ahlrichs, R., Improvements on the direct SCF method. *J. Comp. Chem.* **1989,** *10*, 104.
6. White, C. A.; Johnson, B. G.; Gill, P. M.; Head-Gordon, M., The continuous fast multipole method. *Chem. Phys. Lett.* **1994,** *230*, 8.
7. Strain, M. C.; Scuseria, G. E.; Frisch, M. J., Achieving linear scaling for the electronic quantum Coulomb problem. *Science* **1996,** *271*, 51.
8. Burant, J. C.; Scuseria, G. E.; Frisch, M. J., A linear scaling method for Hartree–Fock exchange calculations of large molecules. *J. Chem. Phys.* **1996,** *105*, 8969.

9. Schwegler, E.; Challacombe, M.; Head-Gordon, M., Linear scaling computation of the Fock matrix. II. Rigorous bounds on exchange integrals and incremental Fock build. *J. Chem. Phys.* **1997,** *106*, 9708.

10. Ufimtsev, I. S.; Martinez, T. J., Quantum chemistry on graphical processing units. 1. Strategies for two-electron integral evaluation. *J. Chem. Theo. Comput.* **2008,** *4*, 222.

11. Ufimtsev, I. S.; Martinez, T. J., Quantum chemistry on graphical processing units. 2. Direct self-consistent-field implementation. *J. Chem. Theo. Comput.* **2009,** *5*, 1004.

12. Rudberg, E.; Rubensson, E. H.; Sałek, P., Kohn− Sham Density Functional Theory Electronic Structure Calculations with Linearly Scaling Computational Time and Memory Usage. *J. Chem. Theo. Comput.* **2011,** *7*, 340.

13. Koch, H.; Christiansen, O.; Kobayashi, R.; Jørgensen, P.; Helgaker, T., A direct atomic orbital driven implementation of the coupled cluster singles and doubles (CCSD) model. *Chem. Phys. Lett.* **1994,** *228*, 233.

14. Hohenstein, E. G.; Luehr, N.; Ufimtsev, I. S.; Martínez, T. J., An atomic orbital-based formulation of the complete active space self-consistent field method on graphical processing units. *J. Chem. Phys.* **2015,** *142*.

15. Vahtras, O.; Almlöf, J.; Feyereisen, M., Integral approximations for LCAO-SCF calculations. *Chem. Phys. Lett.* **1993,** *213*, 514.

16. Beebe, N. H.; Linderberg, J., Simplifications in the generation and transformation of two-electron integrals in molecular calculations. *Int. J. Quantum Chem.* **1977,** *12*, 683.

17. Røeggen, I.; Wisløff-Nilssen, E., On the Beebe-Linderberg two-electron integral approximation. *Chem. Phys. Lett.* **1986,** *132*, 154.

18. Weigend, F., A fully direct RI-HF algorithm: Implementation, optimised auxiliary basis sets, demonstration of accuracy and efficiency. *Phys. Chem. Chem. Phys.* **2002,** *4*, 4285.

19. Aquilante, F.; Pedersen, T. B.; Lindh, R., Low-cost evaluation of the exchange Fock matrix from Cholesky and density fitting representations of the electron repulsion integrals. *J. Chem. Phys.* **2007,** *126*.

20. Koch, H.; Sánchez de Merás, A.; Pedersen, T. B., Reduced scaling in electronic structure calculations using Cholesky decompositions. *J. Chem. Phys.* **2003,** *118*, 9481.

21. Aquilante, F.; Pedersen, T. B., Quartic scaling evaluation of canonical scaled opposite spin second-order Møller–Plesset correlation energy using Cholesky decompositions. *Chem. Phys. Lett.* **2007,** *449*, 354.

22. Pedersen, T. B.; Sánchez de Merás, A. M.; Koch, H., Polarizability and optical rotation calculated from the approximate coupled cluster singles and doubles CC2 linear response theory using Cholesky decompositions. *J. Chem. Phys.* **2004,** *120*, 8887.

23. Friesner, R. A., Solution of the Hartree–Fock equations by a pseudospectral method: Application to diatomic molecules. *J. Chem. Phys.* **1986,** *85*, 1462.

24. Langlois, J. M.; Muller, R. P.; Coley, T. R.; Goddard III, W. A.; Ringnalda, M. N.; Won, Y.; Friesner, R. A., Pseudospectral generalized valence-bond calculations: Application to methylene, ethylene, and silylene. *J. Chem. Phys.* **1990,** *92*, 7488.

25. Ringnalda, M. N.; Belhadj, M.; Friesner, R. A., Pseudospectral Hartree–Fock theory: Applications and algorithmic improvements. *J. Chem. Phys.* **1990,** *93*, 3397.

26. Ko, C.; Malick, D. K.; Braden, D. A.; Friesner, R. A.; Martínez, T. J., Pseudospectral time-dependent density functional theory. *J. Chem. Phys.* **2008,** *128*.

27. Martinez, T. J.; Mehta, A.; Carter, E. A., Pseudospectral full configuration interaction. *J. Chem. Phys.* **1992,** *97*, 1876.

28. Martinez, T. J.; Carter, E. A., Pseudospectral double excitation configuration interaction. *J. Chem. Phys.* **1993,** *98*, 7081.

29. Martinez, T. J.; Carter, E. A., Pseudospectral Møller–Plesset perturbation theory through third order. *J. Chem. Phys.* **1994,** *100*, 3631.

30. Martinez, T. J.; Carter, E. A., Pseudospectral multireference single and double excitation configuration interaction. *J. Chem. Phys.* **1995,** *102*, 7564.

31. Hohenstein, E. G.; Parrish, R. M.; Martínez, T. J., Tensor hypercontraction density fitting. I. Quartic scaling second-and third-order Møller-Plesset perturbation theory. *J. Chem. Phys.* **2012,** *137*.

32. Hohenstein, E. G.; Parrish, R. M.; Sherrill, C. D.; Martínez, T. J., Communication: Tensor hypercontraction. III. Least-squares tensor hypercontraction for the determination of correlated wavefunctions. *J. Chem. Phys.* **2012,** *137*.

33. Parrish, R. M.; Hohenstein, E. G.; Martínez, T. J.; Sherrill, C. D., Tensor hypercontraction. II. Least-squares renormalization. *J. Chem. Phys.* **2012,** *137*.

34. Parrish, R. M.; Hohenstein, E. G.; Schunck, N. F.; Sherrill, C. D.; Martínez, T. J., Exact Tensor Hypercontraction: A Universal Technique for the Resolution of Matrix Elements of Local Finite-Range N-Body Potentials in Many-Body Quantum Problems. *Physical Review Letters* **2013,** *111*, 132505.

35. Song, C.; Martínez, T. J., Atomic orbital-based SOS-MP2 with tensor hypercontraction. I. GPU-based tensor construction and exploiting sparsity. *J. Chem. Phys.* **2016,** *144*.

36. Song, C.; Martínez, T. J., Atomic orbital-based SOS-MP2 with tensor hypercontraction. II. Local tensor hypercontraction. *J. Chem. Phys.* **2017,** *146*.

37. Song, C.; Martínez, T. J., Reduced scaling extended multi-state CASPT2 (XMS-CASPT2) using supporting subspaces and tensor hyper-contraction. *J. Chem. Phys.* **2020,** *152*.

38. Hillers-Bendtsen, A. E.; Mikkelsen, K. V.; Martinez, T. J., Tensor Hypercontraction of Cluster Perturbation Theory: Quartic Scaling Perturbation Series for the Coupled Cluster Singles and Doubles Ground-State Energies. *J. Chem. Theo. Comput.* **2024,** *20*, 1932.

39. Lee, J.; Lin, L.; Head-Gordon, M., Systematically improvable tensor hypercontraction: Interpolative separable density-fitting for molecules applied to exact exchange, second-and third-order Møller–Plesset perturbation theory. *J. Chem. Theo. Comput.* **2019,** *16*, 243.

40. Hillers-Bendtsen, A. E.; Martínez, T. J., Lowering the Scaling of Self-Consistent Field Methods by Combining Tensor Hypercontraction and a Density Difference Ansatz. *J. Phys. Chem. Lett.* **2025,** *16*, 4734.

41. Hillers-Bendtsen, A. E.; Martínez, T. J., Accelerating Hartree–Fock and Density Functional Theory Calculations Using Tensor Hypercontraction. *J. Chem. Theo. Comput.* **2025,** *21*, 11595.

42. Lu, J.; Ying, L., Compression of the electron repulsion integral tensor in tensor hypercontraction format with cubic scaling cost. *J. Comp. Phys.* **2015,** *302*, 329.

43. Lu, J.; Ying, L., Fast algorithm for periodic density fitting for Bloch waves. *arXiv preprint arXiv:1512.00432* **2015**.

44. Hu, W.; Lin, L.; Yang, C., Interpolative separable density fitting decomposition for accelerating hybrid density functional calculations with applications to defects in silicon. *J. Chem. Theo. Comput.* **2017,** *13*, 5420.

45. Yeh, C.-N.; Morales, M. A., Low-scaling algorithm for the random phase approximation using tensor hypercontraction with k-point sampling. *J. Chem. Theo. Comput.* **2023,** *19*, 6197.

46. Parrish, R. M.; Hohenstein, E. G.; Martínez, T. J.; Sherrill, C. D., Discrete variable representation in electronic structure theory: Quadrature grids for least-squares tensor hypercontraction. *J. Chem. Phys.* **2013,** *138*.

47. Kokkila Schumacher, S. I.; Hohenstein, E. G.; Parrish, R. M.; Wang, L.-P.; Martínez, T. J., Tensor hypercontraction second-order Møller–Plesset perturbation theory: Grid optimization and reaction energies. *J. Chem. Theo. Comput.* **2015,** *11*, 3042.

48. Matthews, D. A., Improved grid optimization and fitting in least squares tensor hypercontraction. *J. Chem. Theo. Comput.* **2020,** *16*, 1382.

49. Becke, A. D., A multicenter numerical integration scheme for polyatomic molecules. *J. Chem. Phys.* **1988,** *88*, 2547.

50. Lawson, C. L.; Hanson, R. J., *Solving least squares problems*. SIAM: 1995.

51. Sun, Q.; Berkelbach, T. C.; Blunt, N. S.; Booth, G. H.; Guo, S.; Li, Z.; Liu, J.; McClain, J. D.; Sayfutyarova, E. R.; Sharma, S., PySCF: the Python-based simulations of chemistry framework. *Wiley Interdiscip. Rev. Comput. Mol. Sci.* **2018,** *8*, e1340.

52. Bro, R.; De Jong, S., A fast non-negativity-constrained least squares algorithm. *J. Chemom.* **1997,** *11*, 393.

53. Seritan, S.; Bannwarth, C.; Fales, B. S.; Hohenstein, E. G.; Kokkila-Schumacher, S. I.; Luehr, N.; Snyder, J. W.; Song, C.; Titov, A. V.; Ufimtsev, I. S., TeraChem: Accelerating electronic structure and ab initio molecular dynamics with graphical processing units. *J. Chem. Phys.* **2020,** *152*.

54. Seritan, S.; Bannwarth, C.; Fales, B. S.; Hohenstein, E. G.; Isborn, C. M.; Kokkila-Schumacher, S. I.; Li, X.; Liu, F.; Luehr, N.; Snyder Jr, J. W., TeraChem: A graphical processing unit-accelerated electronic structure package for large-scale ab initio molecular dynamics. *Wiley Interdiscip. Rev. Comput. Mol. Sci.* **2021,** *11*, e1494.

55. Shiozaki, T., BAGEL: brilliantly advanced general electronic-structure library. *Wiley Interdiscip. Rev. Comput. Mol. Sci.* **2018,** *8*, e1331.